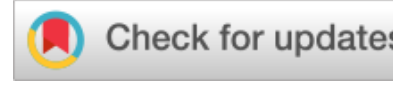

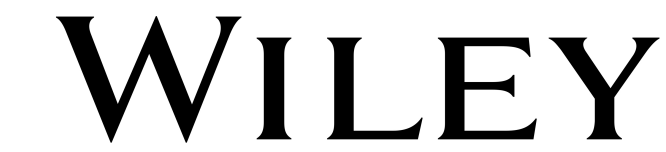

*Research Article*

# Generating Synthetic Light-Adapted Electroretinogram Waveforms Using Artificial Intelligence to Improve Classification of Retinal Conditions in Under-Represented Populations

**Mikhail Kulyabin**,[1] **Aleksei Zhdanov**,[2] **Andreas Maier**,[1] **Lynne Loh**,[3] **Jose J. Estevez**,[3] **and Paul A. Constable**[3]

[1]*Pattern Recognition Lab, Department of Computer Science, Friedrich-Alexander-Universität Erlangen-Nürnberg, Erlangen, Germany*
[2]*Engineering School of Information Technologies, Telecommunications and Control Systems, Ural Federal University, Yekaterinburg, Russia*
[3]*Flinders University, College of Nursing and Health Sciences, Caring Futures Institute, Adelaide, South Australia, Australia*

Correspondence should be addressed to Paul A. Constable; paul.constable@flinders.edu.au

Visual electrophysiology is often used clinically to determine the functional changes associated with retinal or neurological conditions. The full-field flash electroretinogram (ERG) assesses the global contribution of the outer and inner retinal layers initiated by the rods and cone pathways depending on the state of retinal adaptation. Within clinical centers, reference normative data are used to compare clinical cases that may be rare or underpowered within a specific demographic. To bolster either the reference dataset or the case dataset, the application of synthetic ERG waveforms may offer benefits to disease classification and case-control studies. In this study and as a proof of concept, artificial intelligence (AI) to generate synthetic signals using generative adversarial networks is deployed to upscale male participants within an ISCEV reference dataset containing 68 participants, with waveforms from the right and left eye. Random forest classifiers further improved classification for sex within the group from a balanced accuracy of 0.72–0.83 with the added synthetic male waveforms. This is the first study to demonstrate the generation of synthetic ERG waveforms to improve machine learning classification modelling with electroretinogram waveforms.

## 1. Introduction

The full-field flash electroretinogram (ERG) is used as a clinical test of retinal function in disorders affecting this tissue [1]. The guidelines for the clinical ERG testing protocol and calibration of instrumentation are published by the International Society for Clinical Electrophysiology of Vision (ISCEV), which are updated periodically with guidelines for the recording of the ISCEV "standard" ERGs [2, 3]. The standard series of ERG waveforms are recorded in dark-adapted (DA) and light-adapted (LA) conditions to assess primarily the rod and cone photoreceptor pathways of the retina, respectively. The standard series of flash strengths (indicated by the number in $cd{\cdot}s{\cdot}m^{-2}$) include the DA0.01, DA3, DA10, LA3, 30 Hz LA3 flicker, and the filtered DA3 oscillatory potentials (OPs), which together are designed to evaluate different aspects of retinal function [1]. Artificial intelligence (AI) can significantly enhance ERG clinical testing for diseases. The combination of AI and machine learning algorithms can analyze ERG data more accurately and efficiently than traditional methods by detecting subtle patterns and anomalies in the ERG waveforms that might not be immediately apparent to the clinician [4]. The integration and synergistic aspects of AI in ERG testing represent a promising advancement in personalized medicine and ophthalmic diagnostics, given that the ERG can be recorded using noninvasive methods and can provide complementary information to structural, functional, and genetic relationships [5].

Typically, clinical sites performing ERGs would develop their own set of reference datasets with corresponding normative values based on the 95$^{th}$ and 5$^{th}$ centiles for the amplitudes and time to peaks of the principal components of the waveform. However, test conditions such as the electrode position and type and subject parameters such as sex assigned at birth, age, and iris color may also influence the reference ERG parameters [6–8]. Having the ability to increase representative waveforms based on exemplars may help bolster the sample size of not only normative but also of case series in challenging to recruit or under-represented populations in clinical research, such as autism spectrum disorder (ASD) [9–11], rare inherited retinal dystrophies (IRDs) [12], Parkinson's disease [13], glaucoma [14] and attention-deficit/hyperactivity disorder (ADHD) [15].

Recent advances in ERG waveform signal analysis, particularly using features related to the waveform energy and derived from discrete wavelet transform (DWT), have been applied to case-control studies to improve both classification and the understanding of the underlying pathophysiology of retinal disease [14]. This methodological approach was significantly developed by Gauvin and colleagues, who were instrumental in establishing the DWT analysis to the detailed interpretation of the ON- and OFF-signaling pathways in the ERG's a- and b-waves, as well as the higher frequency OPs' components [16–20]. Knowledge of the physiological origins of the energy bands revealed by DWT has helped describe possible differences in the DWT components in neurodevelopmental disorders [10] and in disorders such as glaucoma [14], uveitis [21], and IRDs [22]. Additional signal analytical methods using variable frequency complex demodulation (VFCDM) employs a series of bandpass filters to give a greater time-frequency resolution than DWT, but at the expense of detailed information concerning the cellular origins of the extracted signals [23]. Nonetheless, VFCDM has been applied to the ERG for the classification of ASD and ADHD based on the LA-ERGs [15, 23–25]. In addition to these analytical methods, other developments based on functional data analysis that identify features of the waveform shape to classify groups have also been described for the ascending portion of the b-wave [26].

The application of signal analysis and machine learning to ERGs offers significant potential in classifying complex and rare disorders. This approach is especially valuable for diseases with variable manifestations, such as IRDs, which exhibit a range of phenotypes [12]. Similarly, it can be beneficial in neurodevelopmental and neurodegenerative conditions, where clinical phenotypes are diverse and can be influenced by medication interactions and disease progression. These conditions include ASD and ADHD [9–11, 15, 23, 27] schizophrenia [28, 29], Parkinson's disease [13, 30–32], and Alzheimer's disease [33–35]. The variability and complexity of these disorders make traditional classification challenging, highlighting the need for advanced techniques such as machine learning to improve diagnosis and understanding. This is particularly relevant to rare populations or in conditions with large heterogeneity which require large clinical cohorts to develop robust classification models.

To demonstrate the potential of augmenting machine learning, this report presents the generation of AI-based synthetic ERG waveforms from a dataset containing LA3 and 30 Hz flicker ERGs recorded in healthy volunteers as part of a laboratory reference range. We demonstrate the potential of synthetic ERG synthesis that can support the upscaling of the minority class in a population and improve the overall classification based on sex with the inclusion of synthetically generated male ERGs.

This approach can be applied to any minority class or where additional reference data may be required. The development of synthetic ERGs may also reduce the number of animals used in the studies and improve the statistical accuracy of classification studies in rare and complex populations, further broadening the clinical use of the ERG and its wider diagnostic potential [36, 37] by upscaling existing clinical reference ranges [38, 39] or expanding rare or difficult to reach clinical populations. The implementation of a generative adversarial network approach was used to illustrate the potential of using synthetic data to balance the under-represented male class in a dataset of normal participants.

## 2. Materials and Methods

*2.1. Participants.* In this study, 68 participants, aged between 5.0 and 15.8 years, with no history of ocular or neurodevelopmental disorders, were included. The participants had refractive errors ranging from −6.00D to +2.00D. The group's mean age was 10.4 years (standard deviation (SD): 3.3). Males ($n = 29$) were under-represented in the dataset with ($n = 39$) female participants (68%) forming the majority sex class. The procedures were approved by the Flinders University Human Ethics Committee. The study protocols followed the Declaration of Helsinki. Written informed consent to participate in the study was provided by the parents or guardians of participants, with permission to reuse their data in future studies.

*2.2. Electrophysiology.* ISCEV standard ERGs were recorded using the RETeval (LKC Technologies, Gaithersburg, MD, USA) 6-step ISCEV LA first Troland protocol with adult skin electrodes placed at 2 mm below the lower lid. The Troland protocol used by the RETeval measures the pupil diameter with an infrared camera during recording to maintain a constant retinal illumination for the ERG test [9–11]. This negates the need for dilating eye drops (mydriasis) making the recording more comfortable for the participant. The right eye was recorded first, and after 20 minutes of DA, followed the LA recordings. Signals were filtered by 0.1–300 Hz with automated artifact rejection. The raw-reported averaged waveforms were exported to Excel alongside the main indices for the a- and b-waves time to peak and amplitudes. All recording procedures were in accordance with the manufacturer's recommendations and ISCEV guidelines [2].

For background, the ISCEV standard ERG waveforms are shown in Figure 1. The series consists of a DA series with the initial DA0.01 rod-driven response shown in (a) with a prominent b-wave that originates from rod bipolar cells. At higher flash strengths, the DA3 (d) and DA10 (e) exhibit an a-wave that derives from mainly rod and some cone hyperpolarization with the b-wave following and formed by bipolar, amacrine, and Müller cells in the inner retina. The high-frequency oscillatory potentials (c) are filtered from the DA3 and originate in amacrine cells. Under LA conditions, the transient LA3 response shown in (f) derives from cone pathways with the a-wave formed by hyperpolarization of cones and the b-wave formed by bipolar and amacrine cells with the descending portion of the b-wave also shaped by retinal ganglion cells [2]. The steady state flicker response (c) gives information about the cone function. In this study, we define the time to peak of these waves as "ta" and "tb," respectively, and the amplitudes of these peaks as "la" and "lb" and report the analysis of the LA ERG responses with the full dataset containing DA and LA waveforms.

*2.3. Dataset.* Table 1 provides the number of waveforms from the participants for each stimulus from each eye and the number of samples based on sex. We note the overrepresentation of females in the group. Additional LA3 waveforms were recorded as part of the original study design, and so there are a larger number of LA3 samples in the dataset [40]. The dataset distribution indicates the number of ERG waveforms (including replicates within an eye).

*2.4. Deep Learning Approach*

*2.4.1. Synthesizing ERG Signals.* Generative adversarial networks (GANs) can help deal with unbalanced datasets, a common challenge in many machine learning tasks, including those in medical imaging and time-series domains [41, 42]. Unbalanced datasets occur when some classes have significantly more samples than others, potentially leading to biased models that perform well on the majority class but poorly on the minority classes. The most direct application of GANs, in the context of unbalanced datasets, is their ability to generate synthetic samples for the underrepresented classes. By training a GAN with the condition to generate data belonging to the minority class, it is possible to augment the dataset in a targeted manner, thereby balancing the class distribution without collecting new real-world data, which can be expensive, time-consuming, or impractical. The dataset contains predominantly female individuals. Consequently, generating synthetic male signals was relevant for this demonstration. The conditional GAN (CGAN) to synthesize the synthetic from the natural ERG waveform signals was used in this approach. CGANs are an extension of the GAN framework, designed to generate data conditioned on specific inputs, such as the individual's sex. This conditioning allows CGANs to generate more specific and relevant data for the context, which are particularly useful in the field of visual electrophysiology where retinal signals are used to diagnose and classify a range of conditions using different stimulus parameters [1].

The CGAN architecture comprised two subnetworks: generator and discriminator. The goal of the generator is to learn the transformation between the latent distribution and the real-world data distribution, while the discriminator learns to distinguish the real signals from the synthesized ones. In a CGAN, the generator and the discriminator receive additional conditioning inputs, influencing the data generation process.

For the LA3 and 30 Hz ERG waveforms, the pipeline for synthetic waveform generation using CGAN is shown in Figure 2. The ERG dataset waveforms (LA3 and 30 Hz) were initially split with 20% in the test subset. The CGAN then generated synthetic data equivalent to 20% of the initial real dataset size, specifically for the waveforms from the underrepresented male participants. This step then balanced the sex distribution between males and females in the training data sample following the upscaling of the male sample size. This approach addressed the challenge of generating high-quality synthetic signals with GANs when constrained by a small training dataset.

*2.5. Fourier Decomposition.* We implemented a Fourier decomposition as a standard postprocessing step to mitigate against the effects of overfitting and improve the fidelity of the signals. This step smoothed out noise and peaks in the generated signals, enhancing their overall quality. Fourier decomposition transformed the signals into their frequency components, enabling the selective removal of high-frequency noise that contributed to nonphysiological artifacts. These high-frequency components were attenuated by applying a low-pass filter during this process, which smoothed the signal without losing critical information. This universal postprocessing step was crucial because it compensated for the limited diversity in the training data and ensured that the generated signals were more representative of the real-world initial dataset. The smoothed signals exhibited fewer artifacts, making them suitable for subsequent analyses and applications across various domains such as time-series and time-frequency analyses.

## 3. Results

Figure 3 presents the distributions obtained through t-distributed stochastic neighbor embedding (t-SNE) for the LA3 and 30 Hz datasets for real (male and female) and synthetic (male) datasets. t-SNE is a sophisticated machine-learning algorithm designed to visualize high-dimensional data. It operates by converting similarities between data points into joint probabilities and then minimizing the Kullback–Leibler divergence between the joint probabilities of the high-dimensional data and the corresponding low-dimensional representation. This process allows t-SNE to effectively map complex datasets to a lower-dimensional space, facilitating the identification of patterns, clusters, and relationships that may not be apparent in the original high-dimensional space. These distributions are derived from the values of the ERG signal features la, ta, lb, and tb.

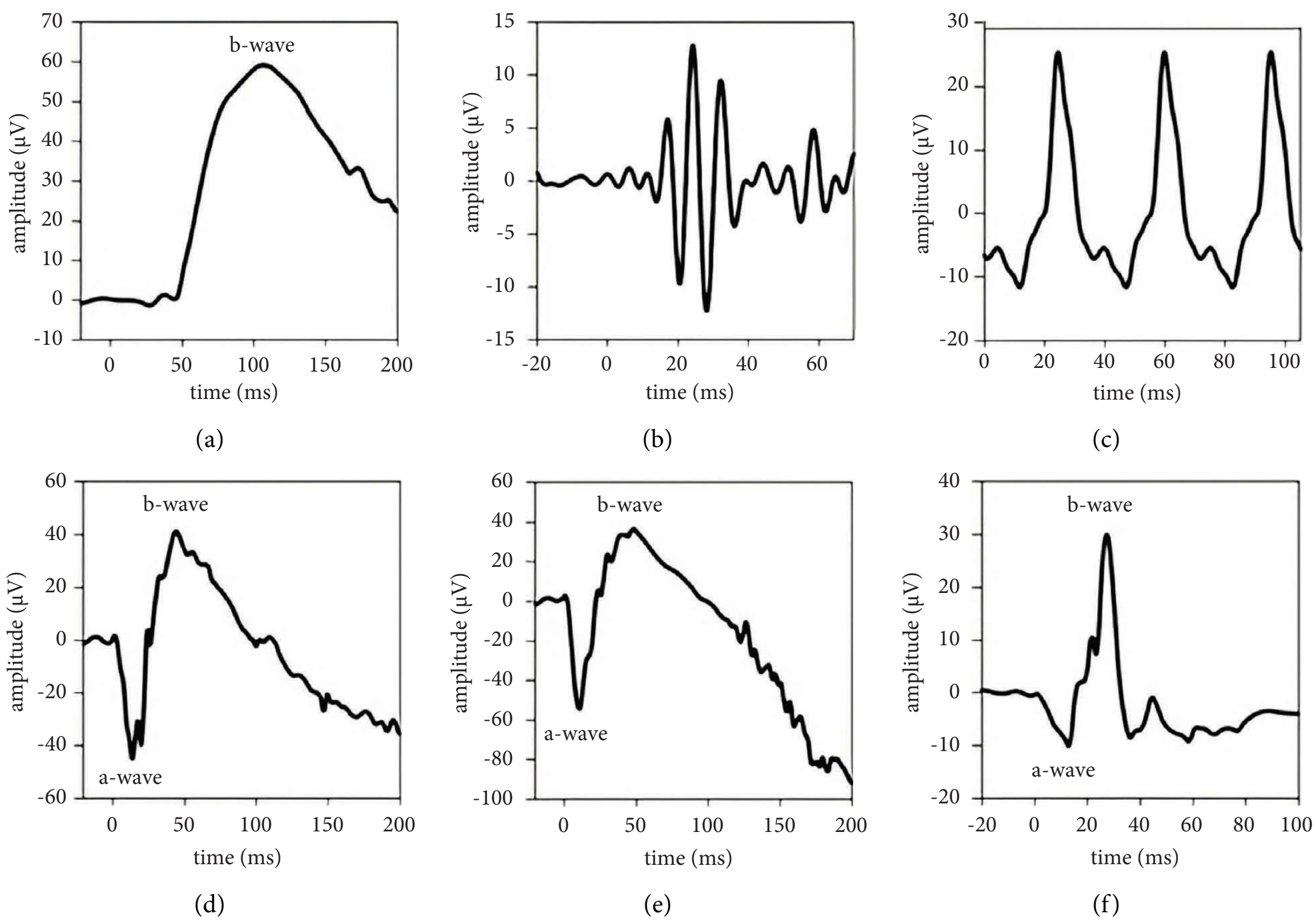

Figure 1: Representative recordings of the ISCEV standard ERGs were recorded with the six-step LA first Troland protocol. The DA0.01 (a), DA3 (d), and DA10 (e) with the DA3 OPs (b) were recorded after 20 minutes of dark adaption. The shape of the ERG waveform varies with the state of retinal adaptation. The LA3 (f) and 30 Hz steady state flicker response (c) indicate retinal cone structure and function. Under DA conditions, the waveform is driven by rod pathways, and under LA conditions, the waveform is driven by the cone pathways. The OPs are filtered from the raw signal to reveal the high-frequency components that originate in amacrine cells in the inner retina. Note different y-axis (amplitude) scales.

Table 1: Number of real ERG recordings from the right and left eyes of the included participants at each of the LA3 and 30 Hz flicker ISCEV standard recordings. Replicates were included in the real dataset to generate synthetic waveforms with males treated as the minority class.

| | LA3 | 30 Hz |
|---|---|---|
| Left eye | 85 | 33 |
| Right eye | 85 | 34 |
| Male ($n = 29$) | 60 | 18 |
| Female ($n = 39$) | 110 | 49 |
| Total waveforms | 170 | 67 |

*3.1. Synthetic Waveforms.* For the analysis of synthetic signals, where the parameters la, lb, ta, and tb were not predefined, the "find peaks" function from the SciPy library package was used. This enabled the identification of these time-domain parameters by detecting peak values within the signal data and slicing them, accordingly, thereby providing additional time-domain features based on the synthetic ERGs for classification modelling based on sex. Figure 4 illustrates the use of SciPy to identify the main a-wave and b-wave peaks in a synthetic LA3 waveform.

Figure 5 shows the representative traces of a real male and female LA3 and 30 Hz flicker ERG waveform with synthetic male waveform for comparison. The main features of the LA3 synthetic male reproduce the timing and morphology of the a-wave, b-wave, and evidence of the OPs on the ascending limb. Similarly, the 30 Hz waveforms are compared with the amplitude and phase of the 30 Hz synthetic waveform aligning with the real waveform. The use of CGAN to generate synthetic representations of real-world data could enable the upscaling of datasets where subjects are rare.

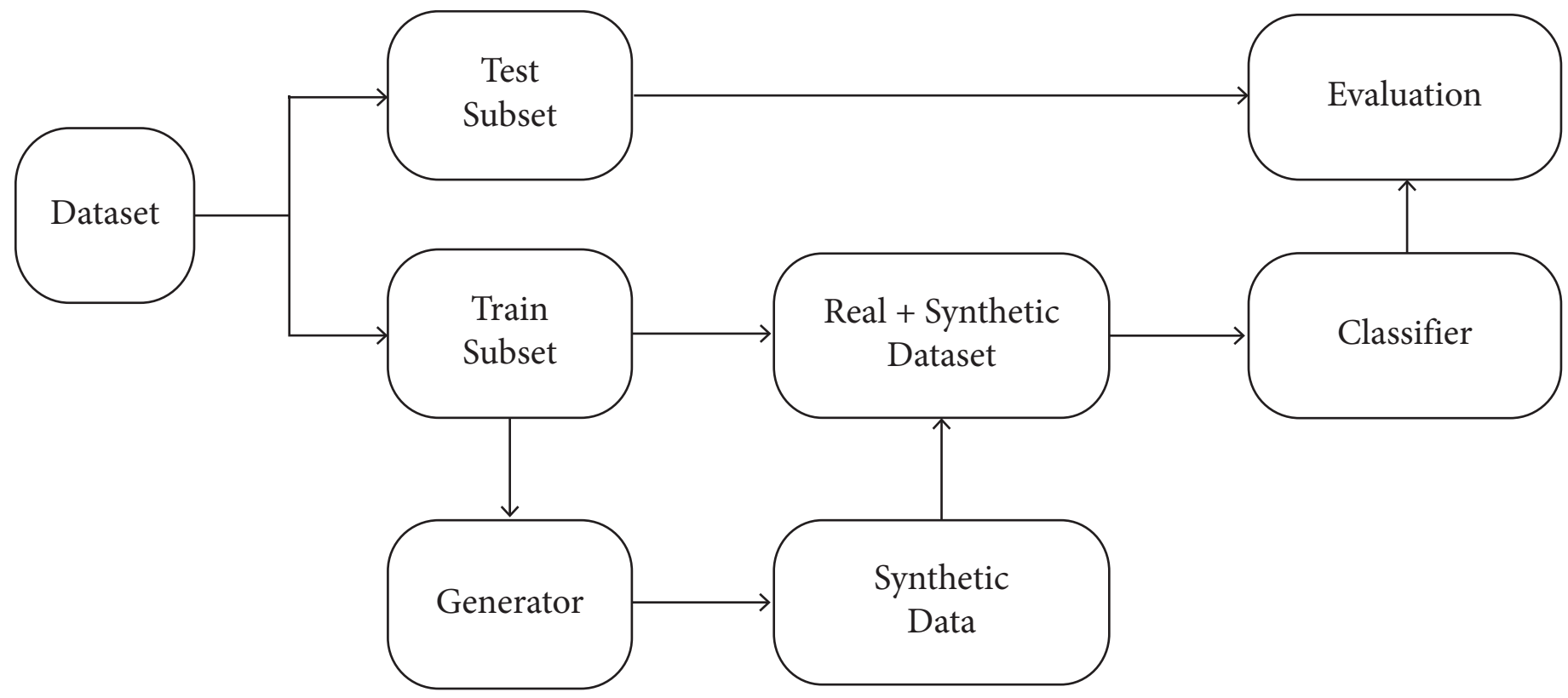

FIGURE 2: Method pipeline. The dataset was split into training (Train) and test subsets. The synthetic signal generator (CGAN) was trained on the Train subset; then, the synthesized signals were added to the real ones from the Train subset, increasing the number of samples from the minority "male" class. Next, a classifier was trained on the augmented and balanced subsets, whose evaluation metrics were computed on the original unbalanced test subset.

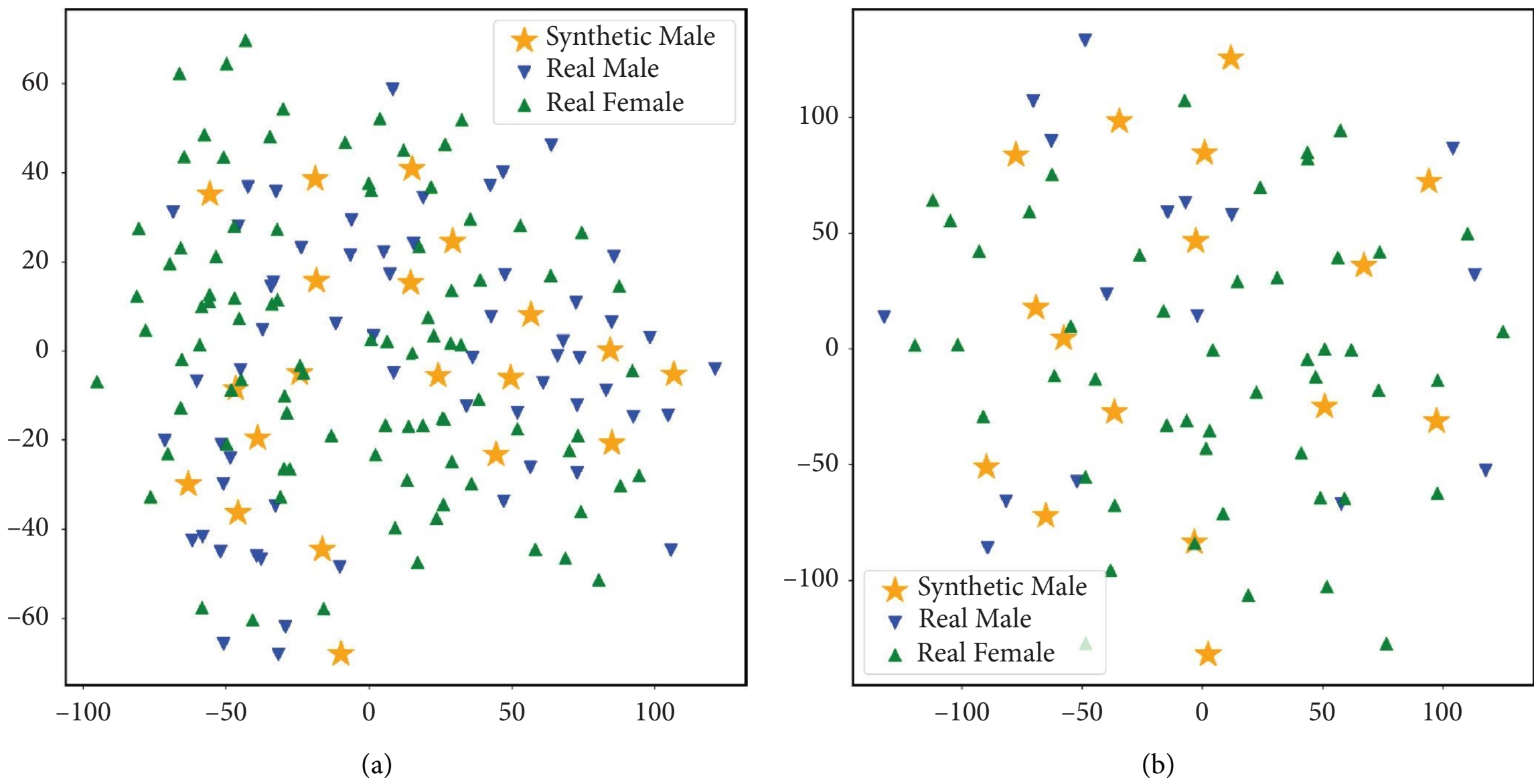

FIGURE 3: t-SNE projections of synthetic male, natural male, and female LA3 (a) and the 30 Hz flicker (b) ERG waveforms.

*3.2. Classification.* The random forest classifier (RFC) was trained on the four parameters (ta, la, tb, and lb) for the LA3 and lb and tb for the 30 Hz flicker. RFC is an ensemble machine learning model that combines multiple decision trees to enhance predictive accuracy and reduce overfitting [43]. GridSearchCV and StratifiedKFold were used with three folds for training and splitting the training set into training and validation subsets with a ratio of 80 : 20. As an alternative, to solve the unbalanced problem of sex within the dataset, an oversampling technique was used to compare with the classification performance with the synthesized data. Here, oversampling was performed as upsampling of the data related to the minority class (male) using replacement.

The balanced accuracy (BA) and *F*1 scores were the highest for the dataset augmented with synthesized data, where BA and *F*1 scores were defined as follows:

$$F1\ \text{score} = \frac{2 \times \text{precision} \times \text{recall}}{\text{precision} + \text{recall}},$$
$$\text{balanced accuracy} = \frac{\text{sensitivity} + \text{specificity}}{2}, \tag{1}$$

where

$$\text{precision} = \frac{\text{TP}}{\text{TP} + \text{FP}},$$
$$\text{recall} = \text{sensitivity} = \frac{\text{TP}}{\text{TP} + \text{FN}}, \tag{2}$$
$$\text{specificity} = \frac{\text{TN}}{\text{TN} + \text{FP}},$$

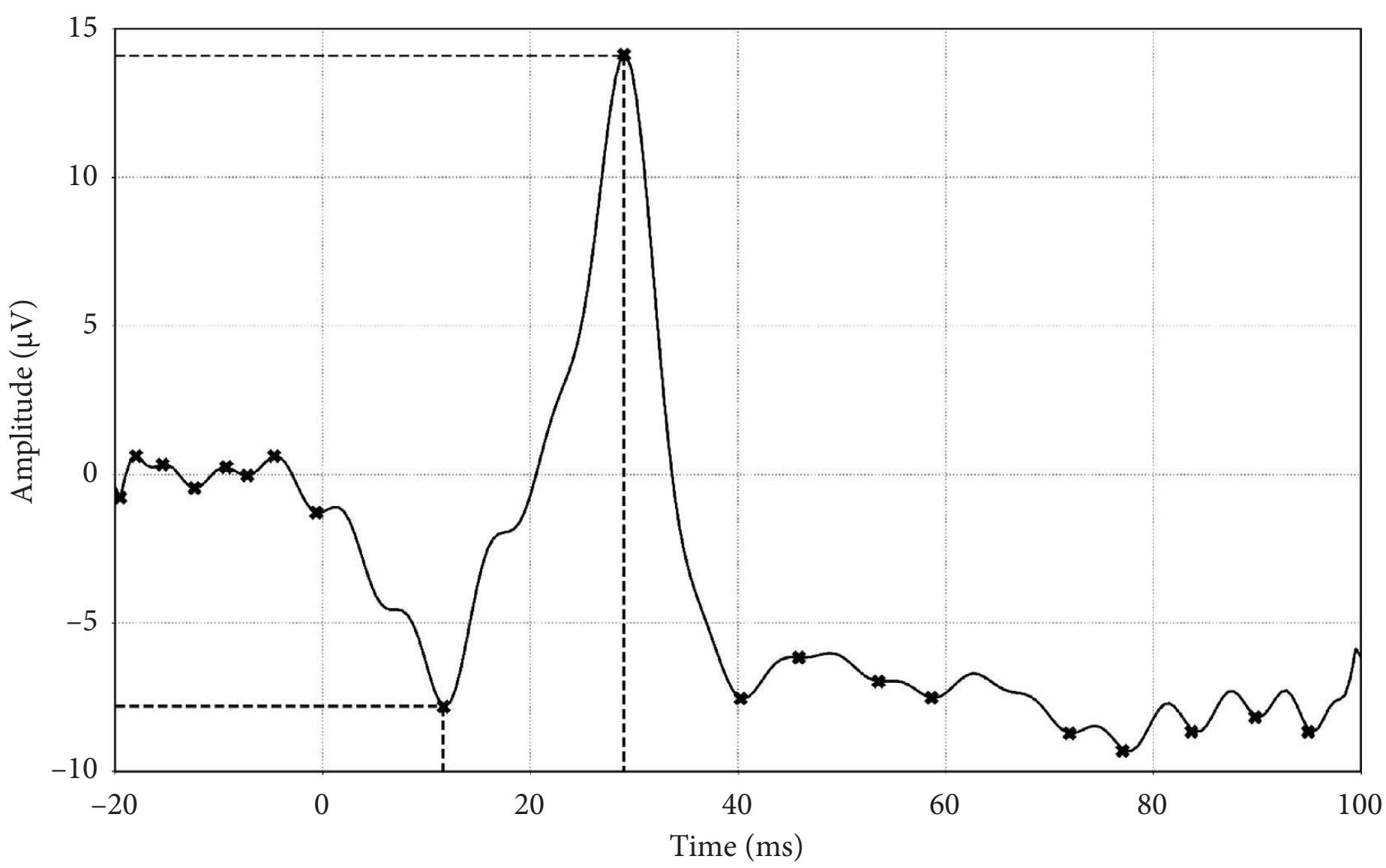

Figure 4: A synthetic LA3 waveform with the la, ta, lb, and tb marked as the respective minima and maxima of the waveform identified using the SciPy library.

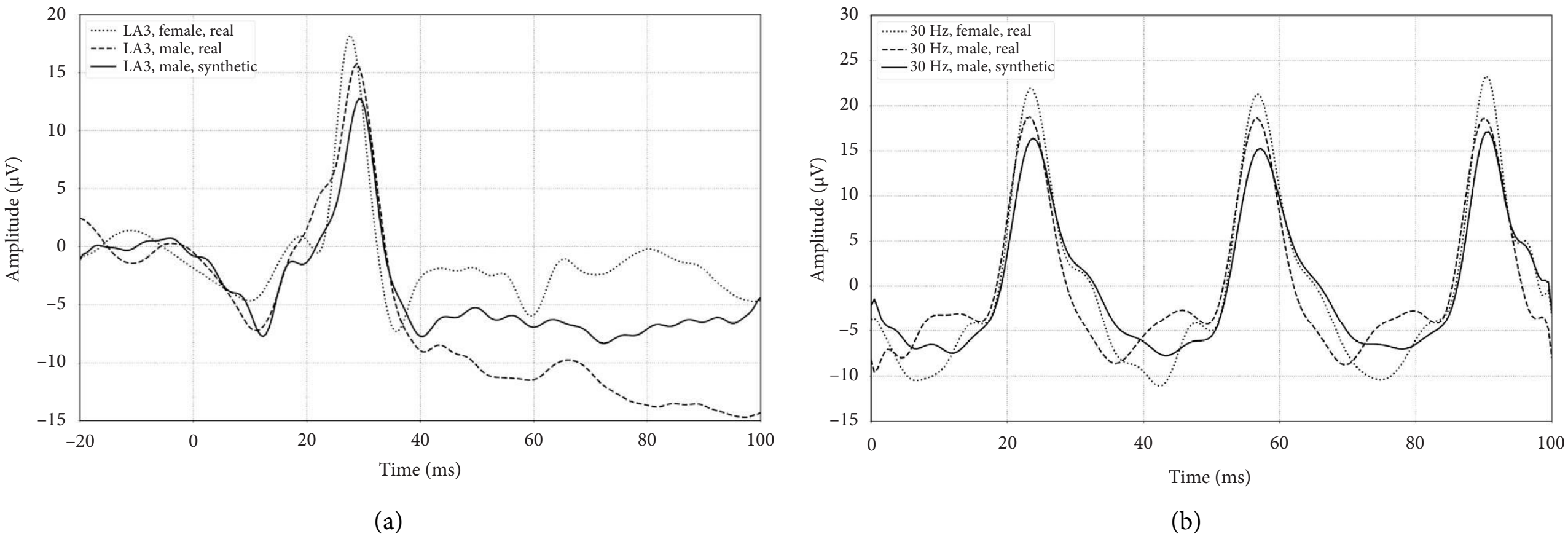

Figure 5: Synthetic male compared with natural male and female LA3 ERG waveform. The LA3 single flash electroretinogram waveform (a) showing a natural male and female response with the synthetic male AI-generated waveform. The a-wave, b-wave, and oscillatory potentials on the ascending limb of the b-wave are presented in the synthetic waveform consistent with the natural waveforms. The 30 Hz flicker electroretinogram waveform (b) showing a natural male and female response with the synthetic male AI-generated waveform. The amplitude and phase of the synthetic waveform are consistent with the natural waveforms.

with TP is true positive, TN is true negative, FP is false positive, and FN is false negative.

Tables 2 and 3 report the random forest classification metrics, balanced accuracy (BA), precision (P), recall (R), and the $F1$ score for the a- and b-wave amplitudes and time metrics for the original, oversampled, and synthesized datasets with superior overall BA obtained with the synthesized dataset.

## 4. Discussion

This report is the first demonstration of improving classification with augmentation of the minority class with synthetic ERG waveform generation. The use of CGAN, in this case, increased the number of male samples for the LA3 and 30 Hz flicker waveforms. This synthetic ERG generation may support future studies where there is a limited number of participants with one or more clinical or sociodemographic characteristics. The features of the synthetically derived waveforms closely resembled those of the real natural time-series data with the a-wave, OPs, and b-wave. Synthetic waveform generation could be applied to other ERG signals such as the pattern ERG [44], the DA cone response (x-wave) [45], multifocal ERG [46] and potentially cortical responses [47].

The generation of additional synthetic waveforms could help with the development of classification models for complex and heterogeneous ocular and neurological conditions. In this first relatively simple case, we showed a proof-of-concept of upsampling the minority class (male) in a sample population and demonstrated an improved classification with the combined synthetic and upscaled dataset. The implications of the use of AI and machine learning may support clinical trials where groups may be hard to access due to geographical isolation or occur with low frequency, improving equity amongst under-represented populations such as Indigenous groups [48]. With expanded test procedures to evaluate endpoints in clinical trials with electrophysiology, psychophysics, and functional tests of vision [49], synthetic ERG waveforms would add to these resources for studies involving rare diseases [50], IRDs [51], or age-related macular degeneartion [52] could also be supported by synthetic ERG waveform generation to improve evaluation of clinical outcomes when clinical datasets are limited.

Table 2: Evaluation metrics of the models for 30 Hz waveforms.

| Dataset analysis | BA | *P* | *R* | *F*1 |
|---|---|---|---|---|
| Original | 0.714 | 0.636 | 0.875 | 0.777 |
| Oversampled | 0.734 | 0.676 | 0.855 | 0.787 |
| Synthesized | 0.785 | 0.710 | 0.814 | 0.823 |

Table 3: Evaluation metrics of the models for LA3 waveforms.

| Dataset analysis | BA | *P* | *R* | *F*1 |
|---|---|---|---|---|
| Original | 0.724 | 0.719 | 0.850 | 0.720 |
| Oversampled | 0.775 | 0.720 | 0.900 | 0.799 |
| Synthesized | 0.825 | 0.809 | 0.850 | 0.829 |

ERG testing is becoming increasingly accessible with the advent of handheld portable devices incorporating the standard and extended ISCEV protocols (https://www.iscev.wildapricot.org). For DA and LA series, the generation of additional synthetic waveforms may improve the analysis of the luminance response functions by efficiently upscaling waveforms across multiple flash strengths [53, 54]. This initial demonstration of the CGAN architecture to generate synthetic waveforms with different morphologies (LA3 and 30 Hz Flicker) indicates the potential for this application to be extended to the pattern and multifocal ERGs [44, 46]. In addition, the cortical evoked potentials such as the flash and pattern visually evoked potentials [55] would also be suited for the generation of synthetic signals further broadening the potential of AI-generated functional measures of vision for future studies.

In this example, the sex assigned at birth was used to demonstrate the proof of principle of applying AI-generated synthetic ERG waveforms to enhance the classification accuracy of the minority "male" class. Whilst sex balance is important for studies where the cases may be biased towards males as in ASD [56], this approach could be used to upsample any group to balance the physiological factors such as iris color [8], refractive error (axial length) [57], or age, especially in pediatric populations, where the ERG waveform develops over time and recording in neonates and infants can be challenging [58].

## 5. Limitations

The generation of synthetic waveforms may not be an exact substitute for real natural waveforms derived from the target clinical population. The spectral composition may differ despite having a subjectively similar form in the time domain. For instance, studies focusing on understanding the generation of ERG b-waves highlight interpretations of physiochemical interactions between retinal cell layers, with frequency ranges varying from fractions of Hertz to several hundred Hertz [59]. Even with natural datasets, the type of instrumentation used to record the waveform may show differences in the spectral composition [60]. Future studies in clinical populations with full waveform decomposition using DWT and/or VFCDM could demonstrate the applicability of AI-generated synthetic ERG waveforms affecting the retina using time-frequency analyses.

The generation of synthetic waveforms through AI is dependent on the size of the dataset used for training, thus necessitating a large original sample. To address this limitation, expanding dataset volumes and promoting open data sharing within the electrophysiology community could enhance the diversity and representation of synthetic waveforms. Although these preliminary results have been generated with a relatively small sample set, sharing ERG datasets between sites would enable larger synthetic datasets to be produced to support clinical studies. One such example is the combination of MRI datasets to support clinical studies as one example that visual electrophysiology clinics could follow [61].

The variability introduced by different recording instruments can affect both natural and synthetic waveforms. Instrumental differences, such as hardware and software filtration in measurement equipment, may neglect high-frequency components in ERG analysis [21]. To mitigate this, normalizing signals considering equipment peculiarities or evaluating result similarities could facilitate data formalization and enhance diagnostic accuracy.

Thus, while AI-generated synthetic ERG waveforms offer promise, their clinical applicability requires further validation in ophthalmology to align with current fields such as cardiology [62]. Future studies employing comprehensive waveform decomposition techniques such as DWT, VFCDM, and vector fitting curve decomposition methods in larger clinical populations could help to elucidate the efficacy and reliability of synthetic waveforms in diagnosing retinal conditions.

## 6. Conclusions

With the expanding clinical utility of the ERG in human and animal studies [63] in fields that extend beyond the retina,

the application of AI-generated synthetic waveforms may benefit future studies in these fields by providing supplementary training data to improve classification models. The findings presented here apply AI-generated synthetic ERG signals to demonstrate the potential for AI to support ophthalmic research into rare, isolated populations or in heterogeneous groups where upsampling of the case or control group is required to balance characteristics to support diagnosis, management, and classification. This would support the expansion and clinical utility of visual electrophysiology [36, 37]. Further work using cortical and macular-derived visual signals would provide an extended capacity to implement CGAN for visual electrophysiological signals.

## Data Availability

The data are available at: Constable, Paul; Kulyabin, Mikhail; Zhdanov, Aleksei E.; Loh, Lynne; Estevez, Jose; Maier, Andreas K. (2024). Generating ISCEV Standard Synthetic ERG Waveforms Using Artificial Intelligence for Enhancing Classification Methods. Flinders University. Dataset. https://doi.org/10.25451/flinders.24893871.v1.

## Ethical Approval

The study was approved by the Flinders University Human Ethics Committee (project 4606). The study protocols followed the Declaration of Helsinki.

## Consent

Written informed consent to participate in the study was provided by the parents or guardians of participants, with permission to reuse their data in future studies.

## Disclosure

This paper has been published as a preprint: Kulyabin, M., Zhdanov, A., Maier, A., Loh, L., Estevez, J. J., and Constable, P. A. (2024). Generating synthetic electroretinogram waveforms using Artificial Intelligence to improve classification of retinal conditions in under-represented populations. arXiv preprint arXiv:2404.11842.

## Conflicts of Interest

The authors declare that they have no conflicts of interest.

## Acknowledgments

The authors thank the participants who took part in this study.